\documentclass[sigconf,natbib=true,anonymous=false]{acmart}
\AtBeginDocument{%
  }

\usepackage{multirow}
\usepackage{booktabs}

\setcopyright{acmlicensed}
\copyrightyear{2018}
\acmYear{2018}
\acmDOI{XXXXXXX.XXXXXXX}
\acmConference[Conference acronym 'XX]{Make sure to enter the correct
  conference title from your rights confirmation email}{June 03--05,
  2018}{Woodstock, NY}
\acmISBN{978-1-4503-XXXX-X/2018/06}

\begin{document}

\title{Not All Transparency Is Equal: Source Presentation Effects on Attention, Interaction, and Persuasion in Conversational Search}

\author{Jiangen He}
\affiliation{%
  \institution{The University of Tennessee}
  \streetaddress{1345 Circle Park Drive}
  \city{Knoxville}
  \state{TN}
  \country{USA}}
\email{jiangen@utk.edu}

\author{Jiqun Liu}
\orcid{0000-0003-3643-2182}
\affiliation{%
  \institution{The University of Oklahoma}
  \streetaddress{401 W Brooks Street}
  \city{Norman}
 \state{OK}
  \country{USA}}
\email{jiqunliu@ou.edu}

\renewcommand{\shortauthors}{He et al.}

\begin{abstract}
Conversational search systems increasingly provide source citations, yet how citation or source presentation formats influence user engagement remains unclear.
We conducted a crowdsourcing user experiment with 394 participants comparing four source presentation designs that varied citation visibility and accessibility: collapsible lists, hover cards, footer lists, and aligned sidebars.
High-visibility interfaces generated substantially more hovering on sources, though clicking remained infrequent across all conditions.
While interface design showed limited effects on user experience and perception measures, it significantly influenced knowledge, interest, and agreement changes.
High-visibility interfaces initially reduced knowledge gain and interest, but these positive effects emerged with increasing source usage. The sidebar condition uniquely increased agreement change.
Our findings demonstrate that source presentation alone may not enhance engagement and can even reduce it when insufficient sources are provided.
\end{abstract}




\maketitle

\section{Introduction}
Conversational search and large-language-model (LLM)–enabled intelligent assistants are reshaping how people seek, judge, and use information. Within the field of human-computer interaction and retrieval (HCIR), decades of research show that interaction design, sense-making, and credible source use are central to effective search~\cite{marchionini2006exploratory, radlinski2017theoretical, zamani2023conversational, liu2021deconstructing}. In particular, user evaluation studies further demonstrate that the interface is not a neutral conduit: presentation choices modulate attention, implicit feedback, and verification behaviors~\cite{joachims2017accurately, hu2022learning, buscher2010good}. In parallel, credibility research in information science establishes that users apply heuristics and strategies sensitive to cues exposed by the interface~\cite{rieh2007credibility, zhang2023design, wang2024cognitively, liu2025boundedly}. For LLM-based systems, transparency features, such as explanations, provenance, and citations, are repeatedly recommended to support scrutiny and appropriate trust~\cite{eiband2018bringing, poursabzi2021manipulating, long2020ai}. Yet, real-world studies of generative search caution that users often do not meaningfully inspect sources even when citations are provided, risking shallow verification, mistrust, and homogenized exposure~\cite{Sharma2024, joseph2025generative, diaz2020evaluating}. Together, these strands indicate that \textit{source-presentation design} is a first-order determinant of credible, accountable conversational and generative interactions.

Despite significant progress, two gaps persist. First, most work examines transparency at a conceptual or system-level (e.g., “provide citations”) rather than as a precise interface variable whose visibility and accessibility can be manipulated and measured. We lack controlled evidence isolating how concrete presentation formats (e.g., inline hover cards vs. footers vs. side panels) affect micro-behaviors such as hovering, clicking, and revisiting sources—the very behaviors through which users enact scrutiny and build credibility judgments~\cite{joachims2017accurately, buscher2010good, huang2011no}. Second, prior studies often stop at global perceptions (e.g., trust, satisfaction, engagement) without connecting interface-induced behaviors to downstream knowledge and attitude change~\cite{yu2025chat}, which are outcomes long recognized in information-seeking and persuasion research~\cite{rieh2007credibility, wang2024cognitively, hilligoss2008developing, draws2021not}. Closing these gaps is crucial for both research and practice: it sharpens theories linking interface affordances, attention, and credibility assessment, and yields prescriptive guidance for designing systems that promote verification without overloading users—particularly vital in domains like health and public policy where misuse of fluent but weakly grounded responses can have real societal costs and diminish public information integrity~\cite{zamani2023conversational, sharma2024generative, dalton2020trec, liu2025boundedly, kim2025medical}.

We address these gaps by experimentally varying how citations are presented during generative search sessions. Specifically, we compare four source-presentation designs that systematically vary \textit{source visibility} and \textit{accessibility}—a collapsible list, inline hover cards, a footer list, and an aligned sticky sidebar—and log their effects on inspection behaviors (hovering/clicking), perceived response quality, and changes in knowledge, interest, and attitude on controversial topics. This design operationalizes interface-level transparency recommendations~\cite{eiband2018bringing} within a conversational and generative IR task framework~\cite{radlinski2017theoretical, alaofi2024generative}, linking interaction traces to cognitive/attitudinal outcomes grounded in credibility and persuasion theory~\cite{rieh2007credibility, petty2012communication}. To address the gaps above, Our crowdsourcing user experiment answers following research questions:

\begin{itemize}
    \item \textbf{RQ1:} How do different source UI designs influence interaction behaviors with source information? 
    \item \textbf{RQ2:} How do source UI designs affect people's perceptions of AI response quality and their experience? 
    \item \textbf{RQ3:} How do source UI designs affect people's knowledge, attitude, and interest change?
\end{itemize}

For RQ1, we examine hovering and clicking patterns across interface conditions to understand how visibility and accessibility affect participants' engagement with sources during information-seeking tasks. To answer RQ2, this study investigates whether different source presentation formats influence perceived credibility, relevance, confidence, satisfaction, and engagement with the AI agent. As a response to RQ3, we explore whether the effectiveness of sources in influencing users' knowledge, interest, and agreement depends on how those sources are presented in the interface.


\section{Methods}









\subsection{Interface Design}
\begin{figure*}
    \centering
    \includegraphics[width=0.8\linewidth]{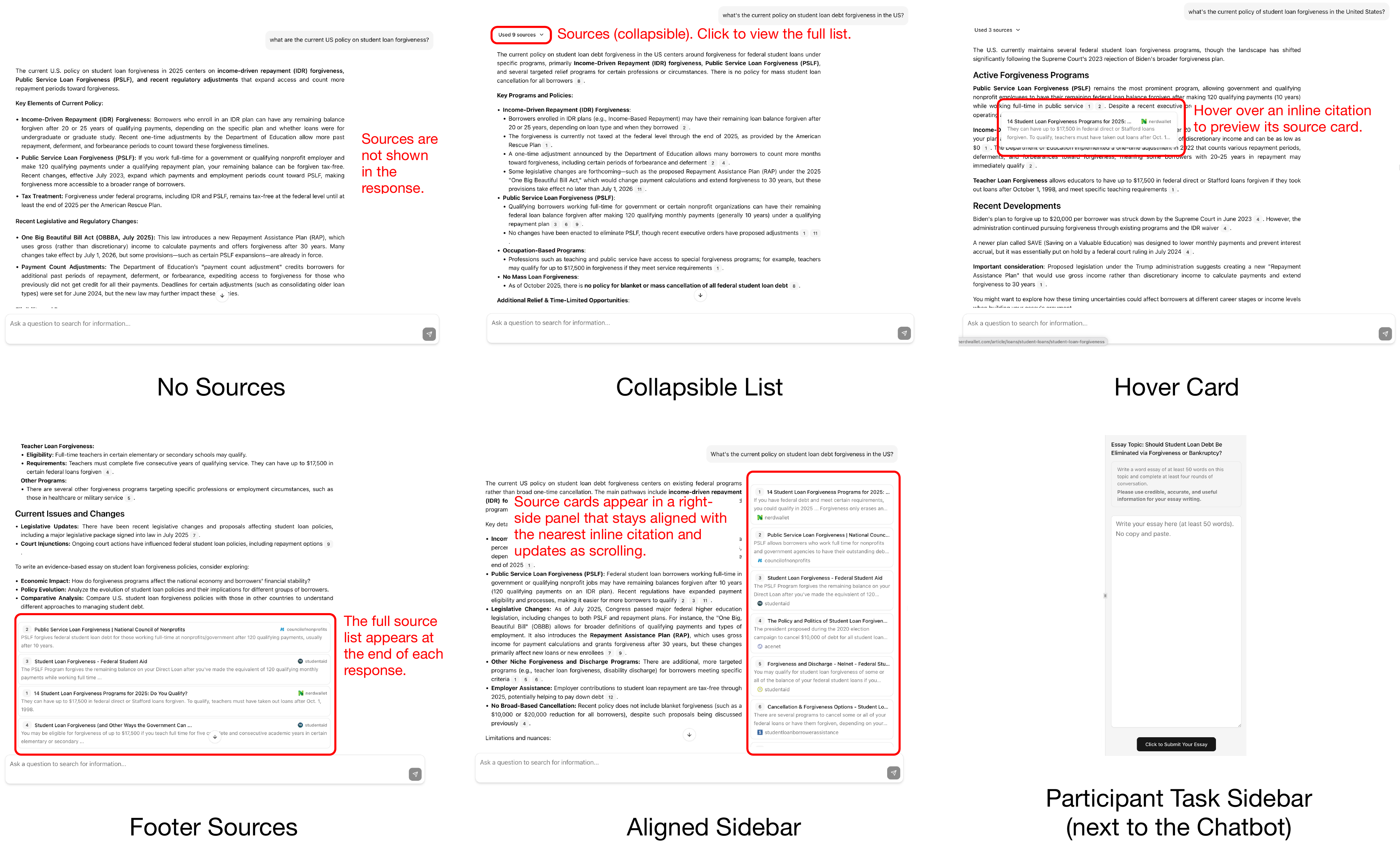}
    \caption{Five user interfaces/conditions.}
    \label{fig:interfaces}
\end{figure*}
To investigate how different source presentation formats influence user experience and chat experience, we implemented five distinct interface conditions (see Figure~\ref{fig:interfaces}).
Except the baseline interface, the four interfaces with source presentation presents the same source details (title, domain name with favicon, and text snippet), but each interface varies in how sources are displayed and accessed. 

\subsubsection{No Sources Interface}

The no sources interface condition serves as a control condition in which no source badges or source cards are shown. Sources are used in generating the responses like the other four conditions, but participants have no access to the underlying sources used to generate the information.

\subsubsection{Collapsible Interface}

In the collapsible interface condition, source badges appear inline within the text as clickable links.
A collapsible ``Used Sources'' button displays at the beginning of each AI response.
Users can click this button to expand and reveal the complete list of source cards containing source details

\subsubsection{Hover Card Interface}

In the hover card interface condition, when users hover over a citation badge, a card appears displaying the source details.
This design provides on-demand access to citation details, allowing users to quickly preview sources while maintaining their position in the text.

\subsubsection{Footer Interface}

In the footer interface condition, all source cards are grouped at the end of each response message. All source cards are displayed by default without the need to expand, compared to the collapsible interface. This approach mimics bibliography formatting, consolidating all reference information in a location. 

\subsubsection{Aligned Sidebar Interface}

The aligned sidebar interface condition presents source cards in a persistent sidebar adjacent to the main content.
Source cards are dynamically positioned to align vertically with their corresponding in-text citations.
The interface employs the bounded isotonic regression to compute optimal non-overlapping positions for the cards. The display of source card updates dynamically in response to scrolling and window resizing.
When citations appear multiple times within the text, positioning is weighted toward the first occurrence using exponential decay to prioritize initial encounters.


We use the four citation interface conditions to manipulate the source visibility and accessibility. Visibility refers to the explicit display of source details, while accessibility refers to the proximity of source details to the in-text citations.
Table~\ref{tab:interface-conditions} summarizes how each interface condition maps to different levels of visibility and accessibility.

The AI agent was powered by Perplexity AI's sonar-pro model, configured with a temperature of 0.7.
The model utilized web search capabilities with medium search context size and user location set to `US' (United States) to retrieve real-time information from the internet and generate responses with inline citations.

\begin{table}[h]
  \caption{Interface Conditions by Visibility and Accessibility}
  \label{tab:interface-conditions}
  \footnotesize
  \begin{tabular}{lcc}
    \toprule
    \multirow{2}{*}{Accessibility} & \multicolumn{2}{c}{Visibility}\\
    \cmidrule(lr){2-3}
                  & Low & High\\
    \midrule
    Low           & Collapsible list & Footer\\
    High          & Hover preview & Aligned Sidebar\\
    \bottomrule
  \end{tabular}
\end{table}
\vspace{-10pt}

\subsection{Experiment Procedure}

The experimental procedure consisted of three stages: a pre-study survey, the main information-seeking and essay-writing task, and a post-study survey. The pre-study survey asked participants to rate their prior experience with and attitudes toward conversational AIs, their confidence in evaluating information, and their intellectual humility~\cite[cf.][]{rieger2024potential}. Participants also rated their familiarity with, interest in, and attitudes toward the controversial topic assigned to them (described below). For the main task, participants were randomly assigned to one of five interface conditions. They received instructions for interacting with the chatbot interface and information about sources, and they were instructed to search for and use credible, accurate, and useful information to write an essay on the assigned topic with at least 50 words. Copying and pasting were allowed during the main task. To encourage richer chat interactions~\cite{liu2025trapped}, we asked participants to engage in at least four turns of conversation. The post-task survey asked participants to rate their perceptions of the AI agent’s credibility, relevance, confidence, satisfaction, response quality, and engagement, as well as their knowledge of, interest in, and attitude toward the topic.

Inspired by the research of \citet{Sharma2024}, we searched for topics in healthcare, science, and news from ProCon.org \footnote{https://www.procon.org/} to engage participants' interest and cognitive load.
We selected three topics:
\begin{enumerate}
\item \textit{News}: ``Should the U.S. government impose stricter regulations on social media platforms to curb misinformation?''
\item \textit{Science}: ``Should the U.S. permit clinical use of germline gene editing in humans?''
\item \textit{Healthcare}: ``Should the U.S. Government Provide Universal Health Care?''
\end{enumerate}

\subsection{Participants}

We recruited 394 participants (195 male, 199 female) with a mean age of 43.63 years (SD = 13.45) on Prolific \footnote{https://www.prolific.co/}.
The inclusion criteria were fluent English-speaking participants from the United States. We also ran evenly split quota sampling for gender and political spectrum (conservative, moderate, liberal).
The median completion time was 17.77 minutes. All the participants were recruited through Prolific and were paid \$3.00 for each session. The study was approved by the Institutional Review Board of the first author's institution.

\section{Results}

\subsection{Behavior Analysis}

\subsubsection{Hovering and Clicking}

\begin{figure}
  \includegraphics[width=1\linewidth]{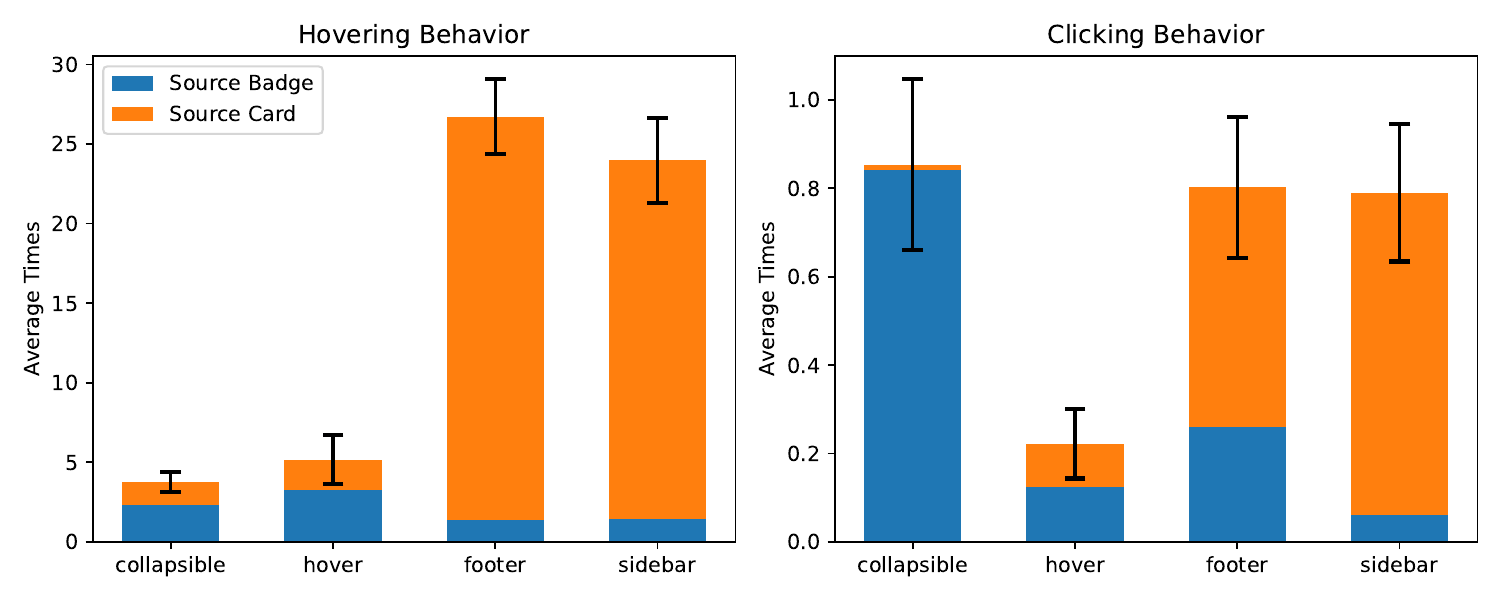}
  \caption{Hovering and clicking by interface condition.}
  \label{fig:hovering_and_clicking_behavior_by_ui}
\end{figure}

We tracked user interactions with source information by logging hover and click events on source cards and citation badges.
A hover event was counted once per continuous hover interaction.
Figure~\ref{fig:hovering_and_clicking_behavior_by_ui} presents the mean hovering and clicking times per participant across the four interfaces.

The footer and sidebar interfaces generated substantially more hovering activity on source cards (26.7 and 24.0 times respectively) compared to the collapsible (3.8 times) and hover card (5.2 times) conditions. It reflects the visibility of source details in source cards in these designs. 
In contrast, clicking behavior is less frequent. Most of participants did not click the sources to read the original sources. Surprisingly, the collapsible interface showed highest click rates (0.85 times) when participants needed to actively expand the source list to access citation details. Conversely, the hover card interface showed the lowest click rates (0.32 times) when participants could preview the source details on hover. The high availability of source details in the hover card interface might lead to participants not clicking the sources to read the sources.

\subsubsection{Participant Prompts and Essay writing}
\begin{table*}[h]
  \caption{Participant Prompts and Essay Writing Statistics (mean and standard deviation) by Interface Condition}
  
  \label{tab:user_message_statistics}
  \footnotesize
  \begin{tabular}{lrrrrrll}
    \toprule
    Metric & No sources & Collapsible & Hover Card & Footer & Sidebar & \multicolumn{2}{c}{Kruskal-Wallis H} \\
    \cmidrule(lr){7-8}
    & & & & & & \textit{H} & \textit{p} \\
    \midrule
    Prompt Count & 4.4 (0.9) & 4.7 (1.4) & 4.8 (1.4) & 4.7 (1.5) & 4.6 (1.1) & 2.269 & 0.519 \\
    Prompt Composition Time (s) & 203.7 (199.6) & 171.7 (179.8) & 165.9 (180.9) & 153.4 (162.6) & 137.8 (153.2) & 4.393 & 0.222 \\
    Prompt Word Count & 72.1 (58.7) & 65.9 (48.3) & 58.7 (36.2) & 61.3 (48.6) & 73.8 (96.8) & 0.170 & 0.982 \\
    Essay Word Count & 97.7 (46.1) & 97.5 (40.8) & 101.1 (47.2) & 89.4 (33.9) & 103.3 (50.7) & 4.705 & 0.195 \\
    Essay Writing Time (s) & 325.6 (222.4) & 311.0 (226.1) & 336.5 (243.1) & 310.9 (188.5) & 340.0 (249.0) & 1.256 & 0.740 \\
    \bottomrule
  \end{tabular}
\end{table*}
Table~\ref{tab:user_message_statistics} presents descriptive statistics for participant prompts and essay characteristics across the five interface conditions.
Kruskal-Wallis H tests revealed no significant differences across interface conditions for any of the metrics.
However, pairwise comparisons using Mann-Whitney U tests revealed significant differences in total message composition time between the no sources condition and several interface conditions with citations. Specifically, the no sources condition differed significantly from the Hover card ($U = 3339.0$, $p = 0.040$), Footer ($U = 3455.0$, $p = 0.013$), and Sidebar ($U = 3747.0$, $p < 0.001$) conditions, with participants in the no sources condition spending more time composing prompts.

\subsection{Perception}
We used ordinal logistic regression models to analyze perception and user-experience outcomes because they were measured on 5-point Likert scales. Each model included the UI conditions (using the no-sources interface as the reference category), an interaction term between UI condition and citation count, demographic variables (age, gender, education), pre-study measures (LLM usage frequency, information-evaluation confidence, intellectual humility, prior knowledge, familiarity, and interest), and behavioral metrics (source hover times, source click times, number of prompts, prompt word count, length of LLM responses, prompt typing time, and total task time). This comprehensive modeling approach allows us to isolate the effects of UI design while accounting for individual differences and engagement patterns.

Table~\ref{tab:regression_results} presents the key regression coefficients for UI conditions and their interactions with citation count across six outcome measures. Coefficients are reported with standard errors and significance levels. Control variables are not displayed in the table but were included in all models. The regression analysis revealed several effects related to UI design and citation presentation. For credibility, a significant Sidebar $\times$ Citations interaction ($\beta = 0.07$, $p < 0.05$) indicated that sidebar presentation becomes more effective with increasing citation counts.
The Footer condition showed a negative main effect on confidence ($\beta = -1.92$, $p < 0.05$) but positive interactions with citation count for both confidence and quality ($\beta = 0.07$ and $0.05$, respectively, $p < 0.05$). No significant effects emerged for relevance, satisfaction, or usefulness.

\begin{table}
  \caption{Regression Analysis Results for Perception Measures}
  \label{tab:regression_results}
  \small
  \begin{tabular}{lcccccc}
    \toprule
    Predictor & Cred & Rel & Conf & Sat & Use & Qual \\
    \midrule
    \multicolumn{7}{l}{\textbf{UI Conditions (Reference: No sources)}} \\
    \quad Collapsible & 0.28 & 1.31 & -1.09 & -0.56 & 0.70 & -1.46 \\
    \quad Hover & 0.49 & 1.01 & -0.86 & 0.51 & -0.47 & -0.71 \\
    \quad Footer & -0.70 & -1.17 & \textbf{-1.92*} & -0.45 & -1.02 & -1.21 \\
    \quad Sidebar & -1.44 & -0.22 & -0.98 & -0.13 & 0.42 & -1.17 \\
    \midrule
    \multicolumn{7}{l}{\textbf{UI × Citations Interactions}} \\
    \quad Colla. × Citations & -0.01 & -0.05 & 0.02 & 0.02 & -0.03 & 0.04 \\
    \quad Hover × Citations & -0.01 & -0.03 & 0.04 & -0.01 & 0.01 & 0.02 \\
    \quad Footer × Citations & 0.03 & 0.04 & \textbf{0.07*} & 0.02 & 0.04 & \textbf{0.05*} \\
    \quad Sidebar × Citations &\textbf{ 0.07* }& 0.02 & 0.03 & 0.02 & -0.01 & 0.07 \\
    
    \bottomrule
  \end{tabular}
  \footnotesize
  Cred = Credibility, Rel = Relevance, Conf = Confidence, Sat = Satisfaction, Use = Usefulness, Qual = Quality. Significance levels: * $p < 0.05$, ** $p < 0.01$, *** $p < 0.001$.
\end{table}

\subsection{Attitude Changes}

Table~\ref{tab:attitude_change_results} presents regression results for three attitude change measures using the same model specification as Table~\ref{tab:regression_results}. The findings reveal more pronounced effects of citation UI design on attitudinal outcomes compared to perception measures. For knowledge gain, both the Footer and Sidebar conditions showed significant negative main effects ($\beta = -1.28$ and $-1.93$, respectively, $p < 0.05$), suggesting reduced self-reported learning when sources were explicitly displayed (high visibility). However, these conditions also exhibited significant positive interactions with citation count ($\beta = 0.05$ and $0.08$, $p < 0.01$ and $p < 0.01$), indicating that their negative impact reversed as citation density increases. 

Interest in the topic was significantly lower across multiple UI conditions: Collapsible ($\beta = -2.45$, $p < 0.01$), Hover ($\beta = -2.84$, $p < 0.01$), and Sidebar ($\beta = -2.69$, $p < 0.01$).
Nevertheless, positive interactions with citation count emerged for Hover, Footer, and Sidebar conditions ($\beta = 0.07$, $0.04$, and $0.07$, respectively, all $p < 0.05$), suggesting that positive effects of source presentation requires sufficient number of citations.

Agreement with the topic claims followed a different pattern: the Sidebar condition increased agreement change ($\beta = 1.89$, $p < 0.05$), although no significant interactions with citation count were observed. This indicates that the positive effect of the Sidebar on agreement change does not require many citations to be presented.

\begin{table}
  \caption{Regression Analysis Results for Attitude Change}
  \label{tab:attitude_change_results}
  \small
  \begin{tabular}{lccc}
    \toprule
    Predictor & Knowledge & Interest  & Agreement \\
    \midrule
    \multicolumn{4}{l}{\textbf{UI Conditions (Reference: No sources)}} \\
    \quad Collapsible & -0.80 & \textbf{-2.45**} & -1.08 \\
    \quad Hover & 0.09 & \textbf{-2.84**} & -1.83 \\
    \quad Footer & \textbf{-1.28*} & -0.98 & 0.59 \\
    \quad Sidebar & \textbf{-1.93*} & \textbf{-2.69**} & \textbf{1.89*} \\
    \midrule
    \multicolumn{4}{l}{\textbf{UI × Citations Interactions}} \\
    \quad Collapsible × Citations & 0.04 & 0.06 & 0.03 \\
    \quad Hover × Citations & 0.00 & \textbf{0.07*} & 0.05 \\
    \quad Footer × Citations & \textbf{0.05**} & \textbf{0.04*} & -0.01 \\
    \quad Sidebar × Citations & \textbf{0.08**} & \textbf{0.07*} & -0.06 \\
    \bottomrule
  \end{tabular}
\end{table}

\section{Conclusion and Future Work}

This study examined how different source citation presentation designs influence user behavior, perception, and attitude change in conversational AI-assisted information seeking.
Through a controlled experiment with five interface conditions (four citation designs that systematically varied visibility and accessibility, plus a no-sources control), we found that interface design significantly affects user engagement with sources and downstream outcomes.
While we observed limited behavioral differences and effects on perception measures such as credibility and satisfaction, the citation presentation format showed more pronounced effects on attitude changes.
Specifically, interfaces with higher citation visibility showed negative main effects on knowledge gain and interest, but these effects were reversed with increasing citation density, suggesting that the effectiveness of transparent source presentation depends on sufficient evidence support.
The Sidebar condition uniquely increased agreement with topic claims, indicating that persistent, spatially aligned citation display may enhance persuasive impact. In future work, we will analyze the content and construction patterns of participant prompts.
A systematic analysis of the essays produced, particularly examining critical analysis and evidence usage, would clarify whether and how citation interfaces influence the quality and rigor of knowledge integration.


\balance
\bibliographystyle{ACM-Reference-Format}
\bibliography{sample-base}










\end{document}